\documentclass[namedreferences]{SolarPhysics}
\usepackage[optionalrh,solaenum]{spr-sola-addons}
\usepackage{graphicx}
\usepackage[usenames]{color}                       
\usepackage{url}                         

\newcommand{\cmt}{cm${}^{-3}$}
\newcommand{\ion}[2]{#1\,{\sc #2}}
\newcommand{\todash}{\,--\,}
\begin{document}

\begin{article}

\begin{opening}

\title{Investigation of Hot X-ray Points (HXPs)  Using Spectroheliograph \ion{Mg}{xii} Experiment Data From CORONAS-F/SPIRIT}

\author{A.~\surname{Reva}$^{1, 2}$\sep
        S.~\surname{Shestov}$^{1}$\sep
        S.~\surname{Bogachev}$^{1}$\sep
        S.~\surname{Kuzin}$^{1}$
       }

\institute{$^{1}$ Lebedev Physical Institute, Russian Academy of Sciences, Leninskii pr. 53, 119991 Moscow, Russia, e-mail:\url{reva.antoine@gmail.com}\\
           $^{2}$ Moscow Institute of Physics and Technology, Dolgoprudnii, 141700 Moscow region, Russia \\
             }

\begin{abstract}
Observations in the  \ion{Mg}{xii} 8.42 \AA\ line onboard the CORONAS-F satellite have revealed compact high temperature objects---hot X-ray points (HXP)---and their major physical parameters were investigated. Time dependencies of temperature, emission measure, intensity, and electron density were measured for 169 HXPs. HXP can be divided into two groups by their temperature variations: those with gradually decreasing temperature and those with rapidly decreasing temperature. HXPs plasma temperatures lie in the range of 5\todash 40~MK, the emission measure is $10^{45}$\todash $10^{48}$~\cmt, and the electron density is above $10^{10}$~\cmt, which exceeds the electron density in the quiet Sun ($10^8$\todash $10^9$~\cmt). HXPs lifetimes vary  between 5\todash 100~minutes, significantly longer than the conductive cooling time. This means that throughout a HXP's lifetime, the energy release process continues, which helps to maintain its high temperature. A HXP's thermal energy is not greater than $10^{28}$~erg, and the total energy, which is released in HXPs, does not exceed $10^{30}$~erg. HXPs differ in their physical properties from other flare-like microevents, such as microflares, X-ray bright points, and nanoflares.
\end{abstract}

\keywords{Corona, Active;  Flares, Microflares and Nanoflares; Spectral Line, Broadening}

\end{opening}

\runningauthor{A. Reva \textit{et al.}}
\runningtitle{Investigation of HXPs Using Spectroheliograph \ion{Mg}{xii} 
Experiment Data}
\section{Introduction}
The process of plasma heating in the solar corona to temperatures beyond 5 MK
occurs due to intense energy release. The study of these processes is
important for understanding the reason for energy release, measuring the
physical conditions in which these processes take place and for compiling
a comprehensive picture of events occurring in the solar corona. Images of
hot plasmas have been obtained from X-ray telescopes such as the
\textit{Yohkoh}/SXT \cite{Tsuneta1991,Ogawara1991}, the
RHESSI \cite{Lin2002}, XRT/\textit{Hinode}
\cite{Kosugi2007,Golub2007},  and the
spectroheliograph~\ion{Mg}{xii}/SPIRIT \cite{zhi03a}. These images showed
that hot plasma is not present everywhere in the solar corona, but just
in its compact areas. High-temperature events of the solar
corona are flares, hot loops, microflares \cite{Lin1984,ben02},
active-region transient brightenings (ARTB: \opencite{shi95}).

Here we present an analysis  of 169 compact high-temperature objects (``hot X-ray points'', HXPs) observed between 20 February 2002 and 28 February 2002 using the spectroheliograph \ion{Mg}{xii} as part of the  CORONAS-F/SPIRIT experiment.

\section{Experimental Data}
\begin{figure}
\centerline{\includegraphics[width= 0.6\textwidth, clip=]{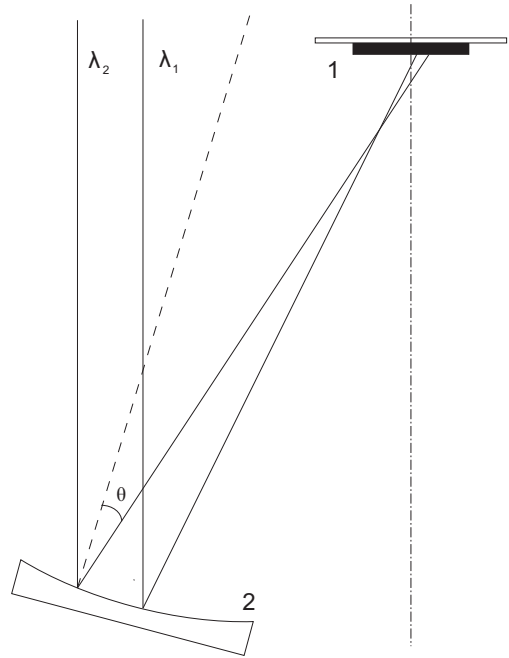}}
    \caption[]{Scheme of the \ion{Mg}{xii} spectroheliograph: 1 -  CCD matrix, 2 - spherical crystal mirror.}
    \label{F:Spectroheliograph_Mg_XII}
\end{figure}

The \ion{Mg}{xii} spectroheliograph is a part of the SPIRIT instrumentation complex, developed in the Lebedev Physical Institute of the Russian Academy of Sciences
\cite{zhi03b}. The \ion{Mg}{xii} spectroheliograph obtains monochromatic images of the solar corona in $\lambda = 8.42$~\AA. A high degree of monochromaticity is achieved by using an optical scheme with a spherical crystal mirror (see Figure~\ref{F:Spectroheliograph_Mg_XII}), for which Bragg's law is satisfied for a narrow wavelength band (only  the \ion{Mg}{xii} 8.42~\AA\ doublet line is detected). Reflection from the surface of the mirror occurs only from its small parts where Bragg's law is satisfied:
\begin{equation}
\label{E:Breg}
2d \cos{\theta} = m \lambda
\end{equation}
(interplanar distance $d$ ($2d = 8.501$~\AA),   angle between incident ray and normal to a mirror $\theta$, order of diffraction $m$, wavelength $\lambda$). For a working wavelength of $\lambda = 8.42$  \AA, Bragg's angle is close to $90^\circ$, and an almost normal incidence occurs. Normal incidence allows us to obtain a high spatial resolution, $\approx 8''$
(effective size of the CCD's pixel is $\approx 6''$).

Equation (\ref{E:Breg}) shows that different wavelengths reflect at different angles, and therefore from different parts of the mirror. Due to spherical aberrations of the mirror, rays that have the same incident angle and a different site of reflection will be focused on different parts of the CCD matrix. Thanks to this effect, the \ion{Mg}{xii} spectroheliograph has a small dispersion and at the same time can build up 2D images.

Due to spectroheliograph dispersion, images in different components of the \ion{Mg}{xii} doublet ($\lambda_1 = 8.4192$~\AA and
$\lambda_2 = 8.4246$~\AA, corresponding to level transitions $1s \ {}^2S_{1/2}$ -- $2p \ {}^2P_{1/2}$ and
$1s \ {}^2S_{1/2}$ -- $2p \ {}^2P_{3/2}$) are shifted from one another. The distance between the two doublet components amounts to five pixels (one pixel is 0.00104~\AA), while the spectral width of the lines amounts to two pixels. An image of the compact source with a size less than one pixel shows a structure that is elongated in the direction of the dispersion.  The transverse width of this structure is defined by a point spread function of the mirror and its FWHM equals $\approx$~one~pixel. Its length amounts to 10--15 pixels and is determined by dispersion of the device and the line's spectral width.

\begin{figure}
\centerline{\includegraphics[width=0.6\textwidth]{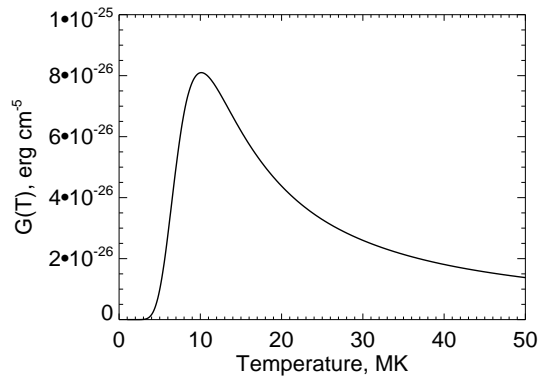}}
    \caption[]{Contribution function of Mg\,{\sc xii} channel}
    \label{F:GOFT}
\end{figure}

\begin{figure}[!h]
\centerline{\includegraphics[width=\textwidth, clip=]{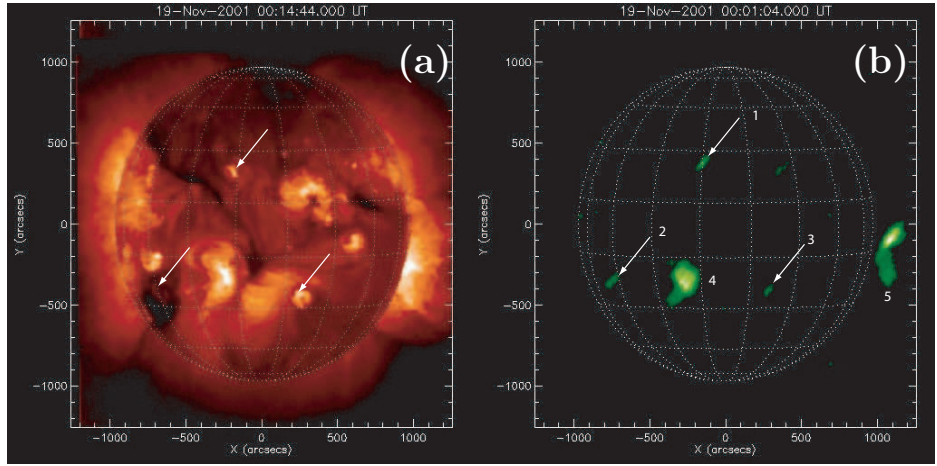}
}
 \vspace{-0.46\textwidth}   
     \centerline{\Large \bf     
      \hspace{0.4 \textwidth}  \color{white}{(a)}
      \hspace{0.4 \textwidth}  \color{white}{(b)}
         \hfill}
     \vspace{0.42\textwidth}    
    \caption[]{Images taken by \textit{Yohkoh}/SXT  (a) and \ion{Mg}{xii} spectroheliograph (b) at close points of time. HXPs are designated by arrows on both images. 1, 2, 3 -- HXP; 4, 5 -- large hot structures.}
    \label{F:compact_structure}
\end{figure}

Emission of the \ion{Mg}{xii}~8.42~\AA\ line occurs in hot plasma with $T>$~5~MK (see Figure~\ref{F:GOFT}). That is why images obtained from \ion{Mg}{xii} spectroheliograph differ from images in cool lines obtained from other telescopes (for example \textit{Yohkoh}/SXT): there is no limb visible on these images and they consist of  separate localized sources. An example of an image obtained using the \ion{Mg}{xii} spectroheliograph is shown on Figure~\ref{F:compact_structure}b. An image obtained from \textit{Yohkoh}/SXT in 2\,--\,40 \AA\ spectral band ($T$~$>$~2~MK) is shown in Figure~\ref{F:compact_structure}a. These two images are taken at close points in time. HXPs are designated with arrows in these figures. Assuming that they are  point sources, the transverse sizes of these structures on the \ion{Mg}{xii} image are determined by spectroheliograph point spread function, and the elongation is caused by spectral dispersion. Images of the \ion{Mg}{xii} spectroheliograph  resemble images from the \textit{Yohkoh}/SXT with Be-filter. HXPs, which are seen on spectroheliograph images, are also seen on Be-filter images.

Observations using the \ion{Mg}{xii} spectroheliograph carried out between 20 and 28 February 2002 were used for this analysis. At that time the satellite was in completely illuminated orbits, and the \ion{Mg}{xii} spectroheliograph obtained images continuously  with cadences from 40\todash 120 seconds. A total of 8689 images were taken.

\section{Temporal Characteristics}
\begin{figure}
\centerline{\includegraphics[width = \textwidth, clip=]{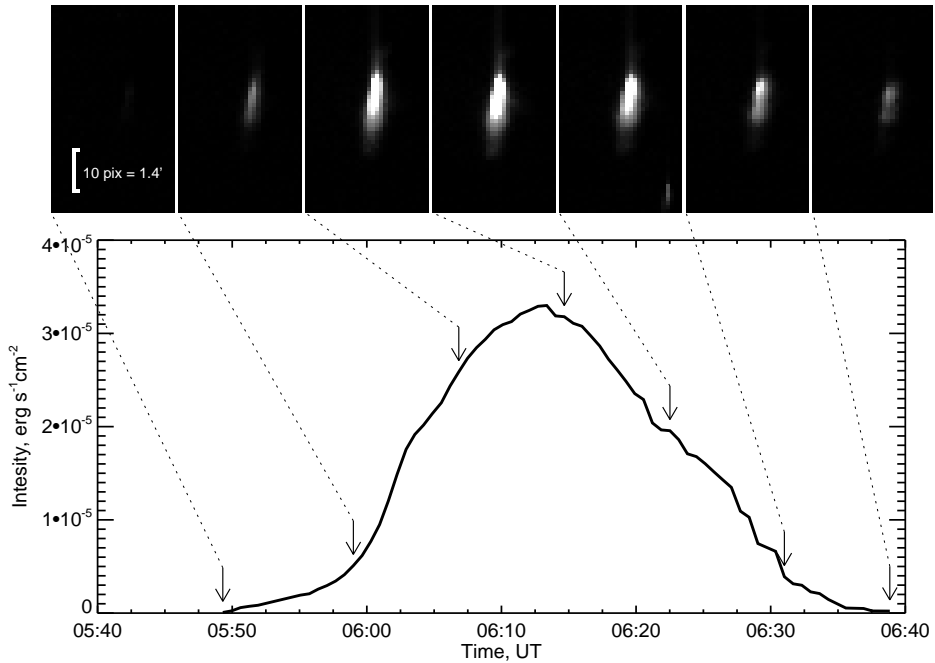}}
    \caption[]{Light curve of a HXP and corresponding date
    images. Times of images are marked on arrows.}
    \label{F:light_curve}
\end{figure}

An example of a series of images with evolving HXP is shown in Figure~\ref{F:light_curve}. On the same image, temporal variations of the intensity (light curve) is also shown. In addition, the times of registration of separate images are marked with arrows. Here, we understand ``intensity of a source'' as a flux from this source in a given spectral band at the Earth's orbit. No  ground-based calibration of the sensitivity of the spectroheliograph was carried out, which is why determination of the sensitivity was carried out using cross-calibration with X-ray data from the GOES satellite \cite{Urnov07}.

\begin{figure}
\centerline{\includegraphics[width = \textwidth, clip=]{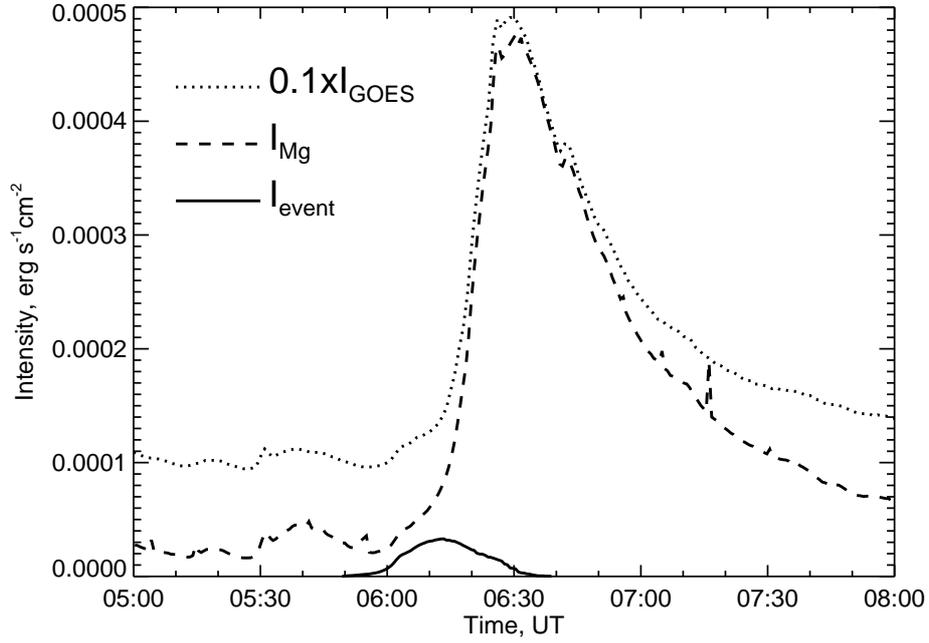}}
    \caption[]{Temporal dependence of HXP intensity  in the 8.42~\AA\ line (solid line), Sun-integrated flux in 8.42~\AA\ line (dashed line) and flux in 1\todash 8~\AA\ GOES channel (dotted line). For clarity, flux in 1\todash 8~\AA\ GOES channel is scaled (multiplied by 0.1)}
    \label{F:goes_mg_hxp}
\end{figure}

Figure.~\ref{F:goes_mg_hxp} shows a light curve of the same HXP, with a temporal dependence of Sun-integrated flux in $\lambda = 8.42$~\AA\ line  and flux in the 1\todash 8~\AA\ GOES channel \cite{Sylwester1995}. For clarity, flux in the 1\todash 8~\AA\ GOES channel is scaled (multiplied by 0.1). This figure shows that the intensity from separate HXP in $\lambda$~=~8.42~\AA\ could amount to 5\% of the total flux from the Sun in $\lambda$~=~8.42~\AA.

\begin{figure}
\centerline{\includegraphics[width = .8\textwidth, clip=]{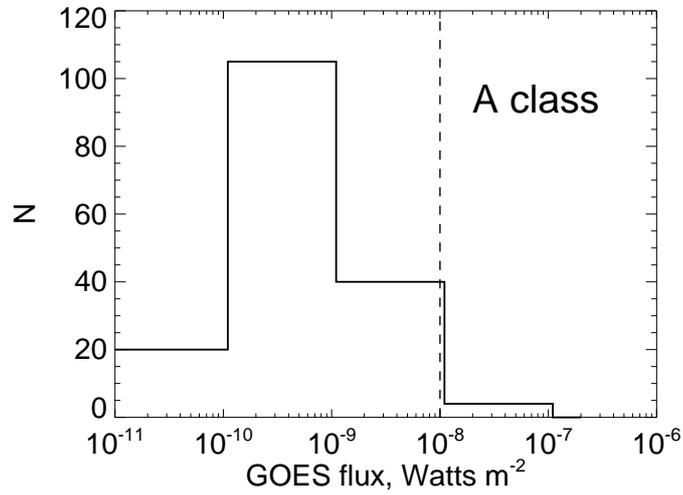}}
    \caption[]{GOES class of HXPs}
    \label{F:GOES_class}
\end{figure}

\inlinecite{Urnov07} showed that there is strong linear correlation between the \ion{Mg}{xii} spectroheliograph  and GOES 1\todash 8~\AA\ fluxes. Peak HXP fluxes were estimated (see~Figure~\ref{F:GOES_class}). Most of the HXPs are below A class.

\begin{figure}
\centerline{\includegraphics[width = \textwidth, clip=]{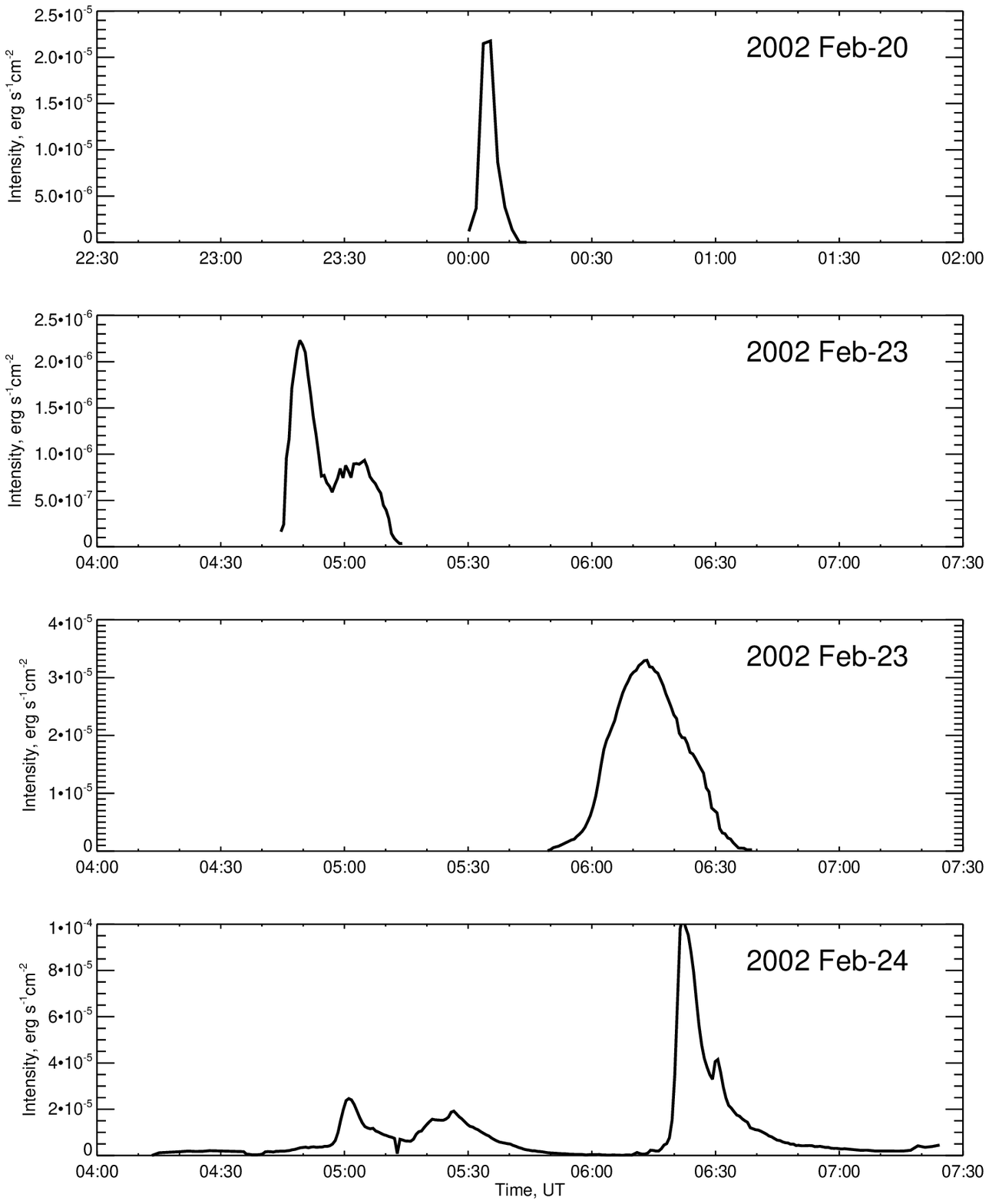}}
    \caption[]{Intensity variation of HXPs}
    \label{F:4_light_curve}
\end{figure}

Examples of a light curve for different HXPs are shown in Figure~\ref{F:4_light_curve}. The lifetimes of these HXPs are approximately 10 minutes, 30 minutes, 50 minutes and 3 hours (we define lifetime as the time between the beginning and the ending of detection of the HXP in the 8.42~\AA\ line). The intensity error  is determined by the noise of the detector at low intensities and by photon statistics at high intensities.

\begin{figure}
\centerline{\includegraphics[width = 0.6\textwidth, clip=]{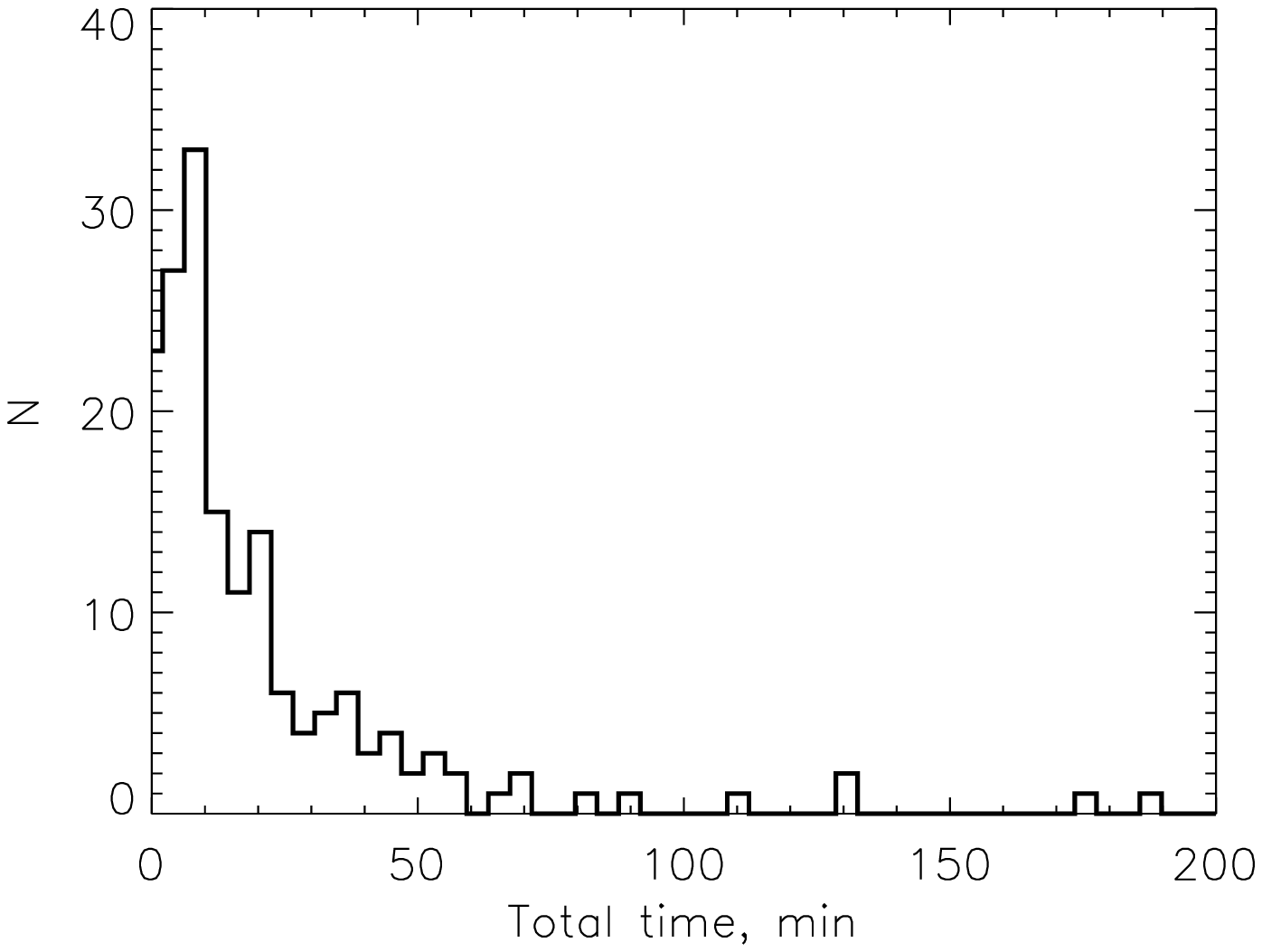}}
    \caption[]{HXP lifetime}
    \label{F:Time_hist}
\end{figure}

A histogram of HXPs' lifetime is shown in Figure~\ref{F:Time_hist}. The most likely lifetime value is ten minutes, although HXPs with lifetime less than two minutes and greater than 60~minutes were observed 20 times, \textit{i.e.} events with very short or very long lifetime are unlikely, but possible.

\section{Determination of HXP Temperature}
The HXP plasma temperature determination is based on measuring the spectral width of the \ion{Mg}{xii} 8.42~\AA\ line. In solar corona conditions the main contribution to line broadenings occurs due to the Doppler effect. Doppler broadening is caused by the thermal and turbulent motion of emitting ions. We neglect turbulence (because estimations show that its influence is small, see Appendix~\ref{A:turb}) and treat HXP plasma as isothermal. Then
\begin{equation}
\frac{\Delta \lambda}{\lambda} = \frac1c \sqrt{\frac{k_B T}{M}} \Rightarrow
T = \frac{c^2 M}{k_B} \left ( \frac{\Delta \lambda}{\lambda} \right)^2,
\label{E:temperature}
\end{equation}
where $\Delta \lambda$ is the Doppler broadening, $\lambda$ = 8.42~\r{A}, $c$ is the speed of light, $k_B$ is Boltzmann's constant, $T$ is the plasma temperature, and $M$ is the mass of a magnesium ion.

\begin{figure}
\centerline{
\includegraphics[width = 0.4\textwidth, clip=]{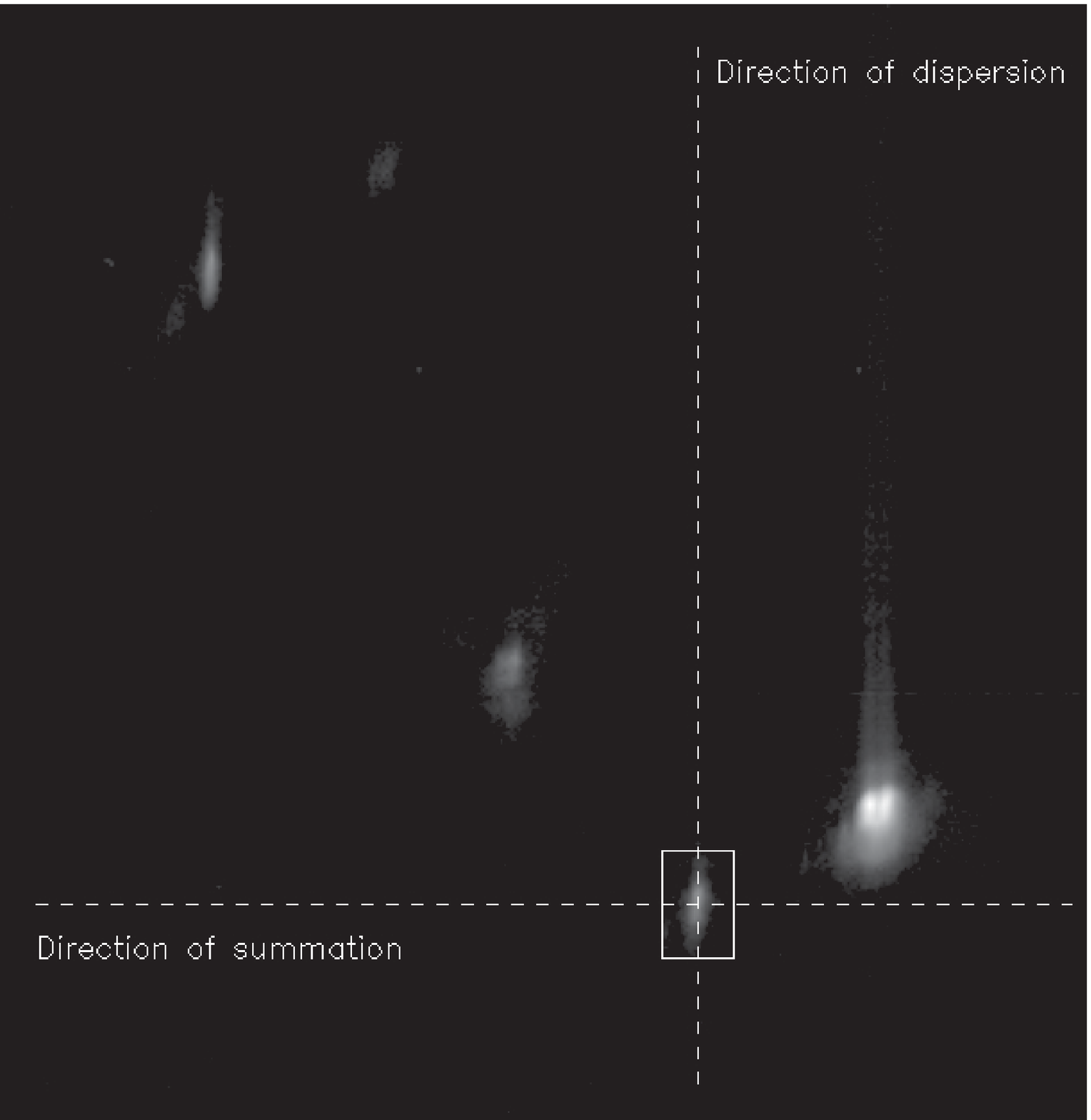}
	 \includegraphics[width = 0.58\textwidth, clip=]{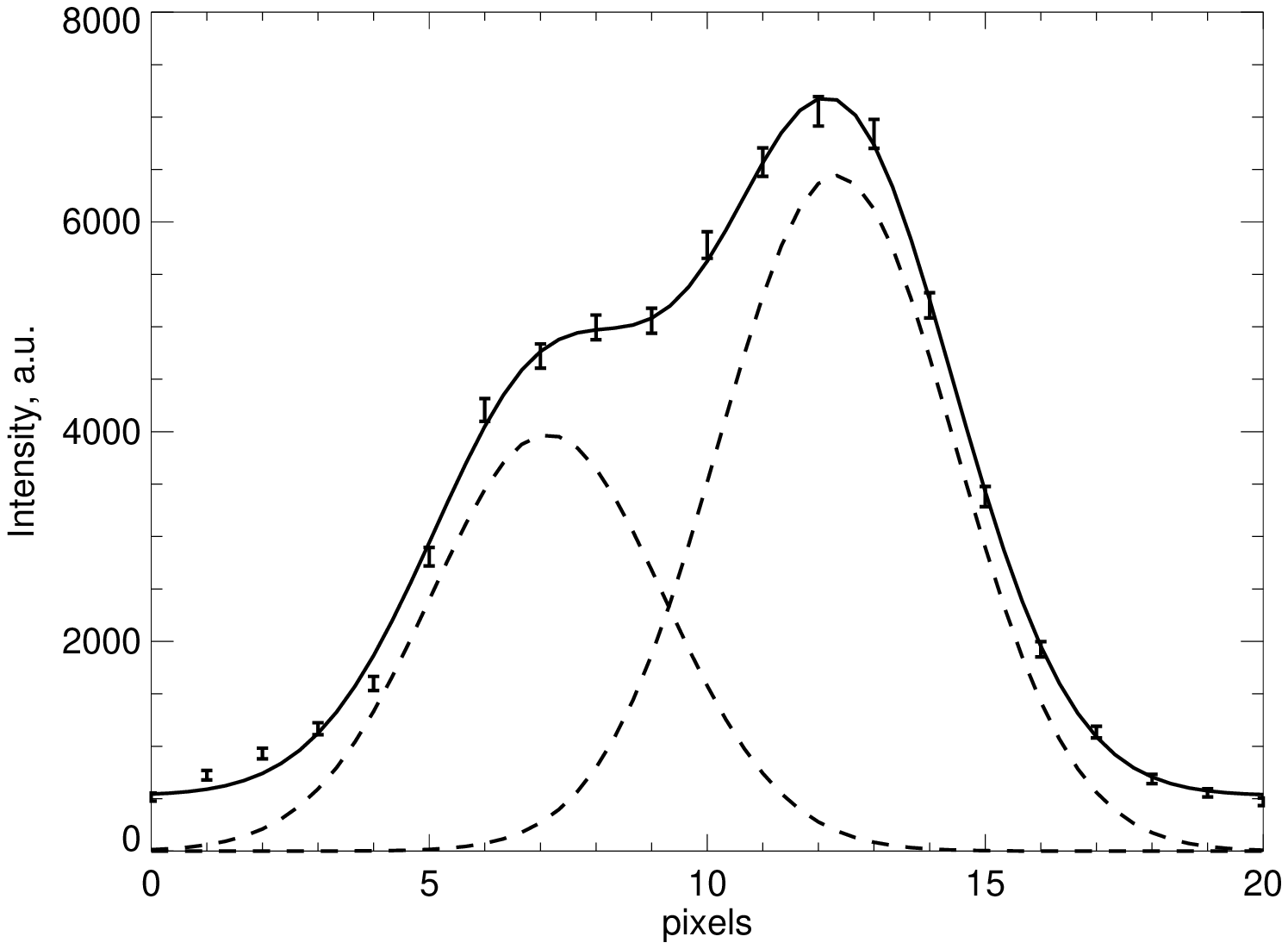}
}
 \vspace{-0.4\textwidth}   
     \centerline{\Large \bf     
      \hspace{0.0 \textwidth}  \color{white}{(a)}
      \hspace{0.41\textwidth}  \color{black}{(b)}
         \hfill}
     \vspace{0.36\textwidth}    
    \caption[]{Magnesium doublet. a) Spectroheliograph image with HXP outlined (direction of dispersion is vertical, direction of summation is horizontal). b) Spectrum obtained after summation.}
    \label{F:spectra_determination}
\end{figure}

Since only objects with small transverse sizes (relative to the direction of dispersion) were selected for analysis, we assume that their longitudinal size is also small. This means that a HXP's intensity profile in the dispersion direction is a spectrum of \ion{Mg}{xii} $\lambda = 8.42$~\AA\ line, undistorted by its spatial structure. Scanning of compact sources (in the dispersion direction) allows us in most cases to resolve both components of the \ion{Mg}{xii} doublet. In order to obtain a doublet spectrum, the spectroheliograph signal  was summed in the vicinity of a HXP in the direction perpendicular to the  dispersion. This method is illustrated on Figure~\ref{F:spectra_determination}.

The spectrum obtained was approximated by the sum of two Gauss profiles:
\begin{equation}
I(\lambda) = I_1\exp\left[-\frac{(\lambda-\lambda_1)^2}{2\Delta\lambda^2}\right]+
I_2\exp\left[-\frac{(\lambda-\lambda_2)^2}{2\Delta\lambda^2}\right]+I_0
\end{equation}
where $I(\lambda)$ is  measured spectrum. $I_1$, $I_2$,  $I_0$, $\lambda_1$, and $\Delta\lambda$ are unknown spectrum parameters, which are determined by  fitting.  The difference between $\lambda_1$ and $\lambda_2$ is fixed and equals 5.4 m\AA. It is worth mentioning that the temperature cannot always be determined: at low intensities, noise of the CCD  corrupts the signal; also, sometimes the transverse size of the HXP could exceed one pixel, and therefore  the temperature using the above method would not be accurate.

\begin{figure}
\centerline{
\includegraphics[width = \textwidth, clip=]{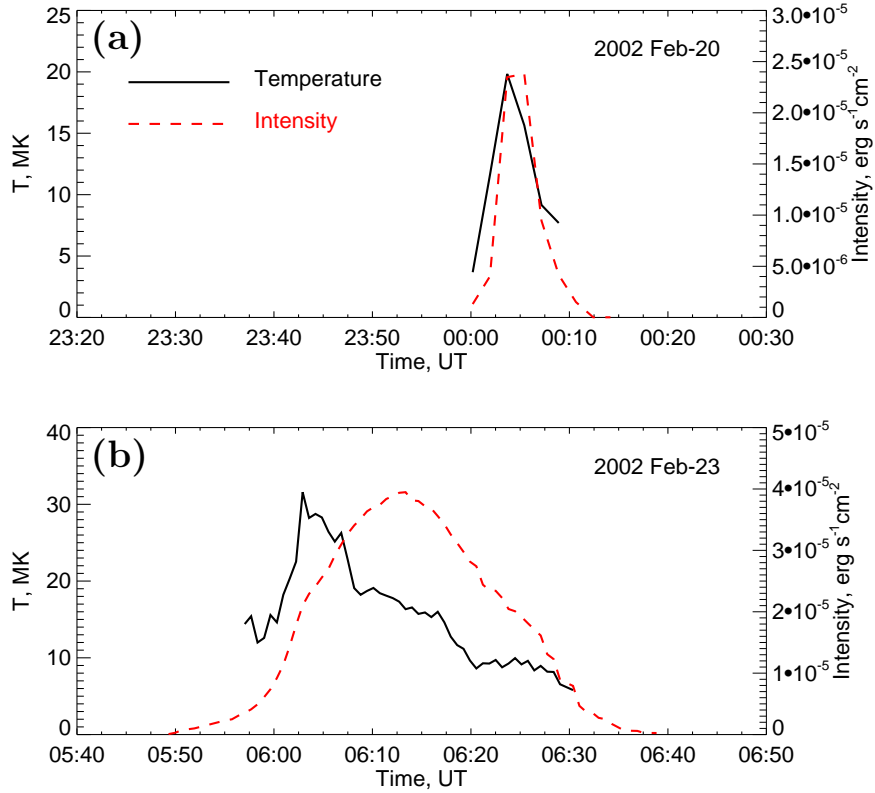}}
 \vspace{-0.86\textwidth}   
     \centerline{\Large \bf     
      \hspace{0.13 \textwidth}  \color{black}{(a)}
      \hfill}
     \vspace{0.41\textwidth}    
       \centerline{\Large \bf     
      \hspace{0.13 \textwidth}  \color{black}{(b)}
      \hfill}
     \vspace{0.35\textwidth}    
    \caption[]{Temporal variations of HXPs' temperature and intensity. Solid line --- temperature, dashed line --- intensity}
    \label{F:Temperature_curve}
\end{figure}

In Figure~\ref{F:Temperature_curve}, the variations of intensity and temperature of two HXPs are shown. The first event is relatively fast. It lasts ten minutes and the temperature reaches 20~MK. Maximum  intensity and temperature occur approximately at the same time. The second HXP lasts 50~minutes, and the temperature reaches 30~MK. The maximum intensity is delayed by 20~minutes relative to maximum temperature. This effect is also seen for a number of other HXPs.

Let us mention that the temperature is always greater than 5~MK, and as the temperature reaches 5~MK, the intensity approaches zero. These two facts are in agreement with the contribution function of the \ion{Mg}{xii} 8.42~\AA\ line. The contribution function decreases rapidly when the temperature approaches 5~MK.

HXPs' peak temperature (that is the maximum temperature  reached during evolution of an individual HXP) lies in the range of 5\todash 50~MK.

\begin{figure}
\centerline{\includegraphics[width = 0.6\textwidth, clip=]{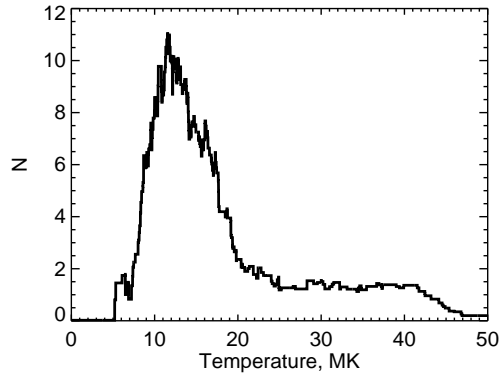}}
    \caption[]{Weighted histogram of HXPs' peak temperature. Each HXP is represented in the figure by a box of unit area, the width of which is $2 \Delta T$.}
    \label{F:Temperature_hist}
\end{figure}

A histogram of the HXP's peak temperature is shown in Figure~\ref{F:Temperature_hist}. The most likely value of the peak temperature is 12~MK.

\begin{figure}
\centerline{\includegraphics[width= 0.6\textwidth, clip=]{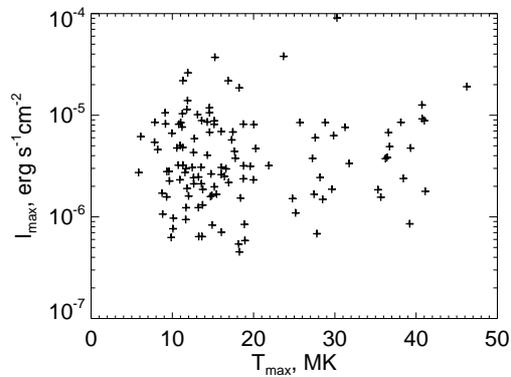}}
    \caption[]{Dependence of HXPs' peak intensity on peak temperature.}
    \label{F:i_max_t_max}
\end{figure}

The dependence of peak intensity on  peak temperature is shown in Figure~\ref{F:i_max_t_max}. This figure shows that the two quantities are not correlated, \textit{i.e.} more intense events are not necessarily hotter ones. Therefore, knowledge of peak intensity or peak temperature is not enough for the description of a HXP, but rather knowledge of other physical parameters is required.

\section{Absolute Intensity, Emission Measure, and Electron Density}

\begin{figure}
\centerline{\includegraphics[width = 0.6\textwidth, clip=]{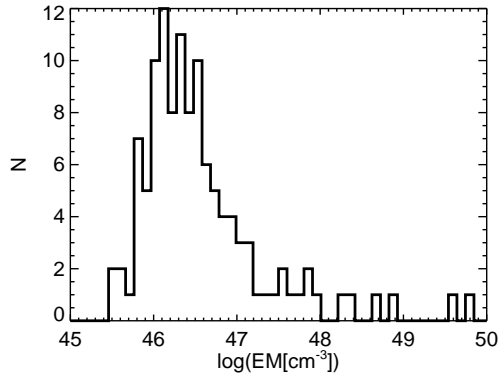}}
    \caption[]{Histogram of HXPs' peak emission measure}
    \label{F:em_hist}
\end{figure}

The HXP's emission measure is calculated from the absolute intensity and temperature. A histogram of HXP's peak emission measure is shown in Figure~\ref{F:em_hist}. The HXP emission measure lies in the range of $10^{45}$--$10^{48}$~\cmt.

If volume and emission of a source are known, its electron density could be determined. The size of an HXP cannot be measured, but we can make an estimation of its upper value (not greater than 5~Mm) and therefore arrive at an estimation of the lower limit of the electron density:
\begin{equation}
\label{e:n_est}
n_e \ge \sqrt{\frac{EM}{h_0^3}}
\end{equation}
This estimation gives us a value of $10^{10}$~\cmt, which is higher than the typical value of electron density in the quiet corona ($10^8$\todash $10^9$~\cmt).

It is worth mentioning that an estimation of the sensitivity of the \ion{Mg}{xii} spectroheliograph from the optical properties of the elements of which it consists (reflection coefficient of mirror on the working wavelength, detector sensitivity, \textit{etc.}) is ten times higher than the value obtained by cross-calibration with GOES data. This means that obtained values of intensity and emission measure are accurate within a  factor of ten. The uncertainty in the electron density would be three, which is acceptable for an estimation.

\section{Dynamics of HXPs}
For all 169 analyzed HXPs temporal variations of intensity, temperature, emission measure, and electron density were obtained. HXPs could be divided into two groups by the behavior of these parameters: those with gradually decreasing temperatures and those with rapidly decreasing temperatures.

\begin{figure}
\centerline{\includegraphics[width = \textwidth, clip=]{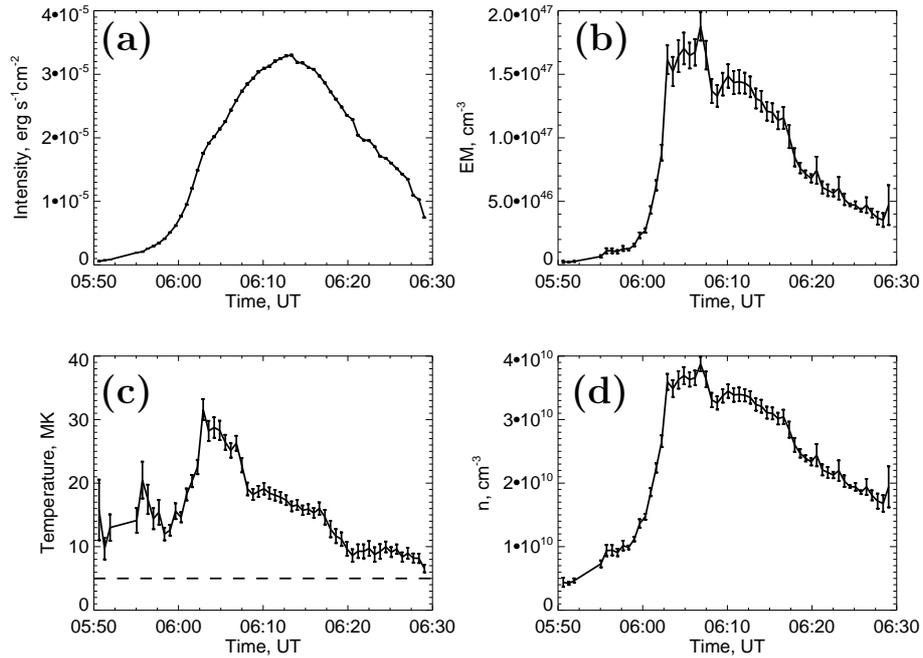}
}
 \vspace{-0.71\textwidth}   
     \centerline{\Large \bf     
      \hspace{0.09 \textwidth}  \color{black}{(a)}
      \hspace{0.42\textwidth}  \color{black}{(b)}
         \hfill}
     \vspace{0.34\textwidth}    
     \centerline{\Large \bf     
      \hspace{0.09 \textwidth}  \color{black}{(c)}
      \hspace{0.42\textwidth}  \color{black}{(d)}
         \hfill}
     \vspace{0.29\textwidth}    
    \caption[]{Variation of intensity (a), emission measure (b), temperature (c), and electron density (d) of an HXP of the first group (23 February 2002).}
    \label{F:T_I_n_EM}
\end{figure}

Let us consider the first group (see Figure~\ref{F:T_I_n_EM}). HXPs of this group  have low intensity and emission measure at the beginning of their lifetime; their temperature lies in the range of 10\todash 15~MK. During a period of approximately five minutes, their intensity and emission measure increase slowly, and the temperature is roughly constant. Then the temperature reaches extremely high values (30\todash 50~MK) over a period of 2\todash 5~minutes, while its emission measure reaches a maximum, and intensity increases with the same rate. After reaching its maximum temperature, the HXP cools down to 5~MK between 5\todash 30~minutes, while the emission measure slowly decreases. The intensity reaches its maximum after maximum temperature. As the temperature approaches 5~MK, the intensity approaches zero. Since the cadence of the spectroheliograph is limited, for some events the phase of temperature increase is not seen. That is why some events are seen beginning from high temperatures (30\todash 50~MK). This behavior of HXP parameters could be explained by the following scenario: at some moment in time, a process of energy release in the volume of HXP begins. As a result, the HXP temperature increases and reaches its maximum. Then the power of the energy release decreases and the HXP starts to slowly cool down. HXPs of this group are 39\% of the whole of observed HXPs.

\begin{figure}
\centerline{ \includegraphics[width = \textwidth, clip=]{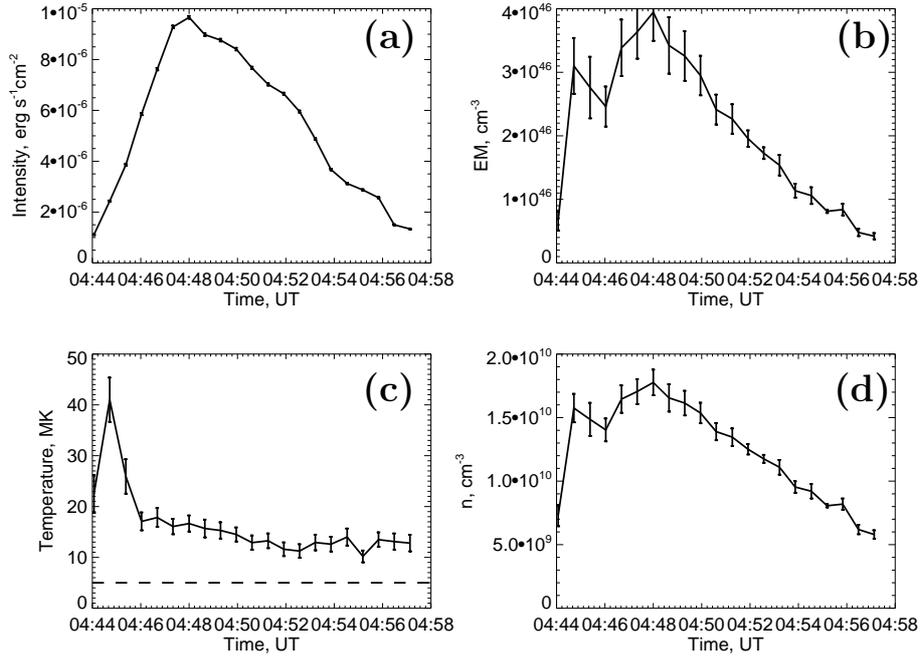}
}
    \vspace{-0.71\textwidth}   
     \centerline{\Large \bf     
      \hspace{0.38 \textwidth}  \color{black}{(a)}
      \hspace{0.42\textwidth}  \color{black}{(b)}
         \hfill}
     \vspace{0.34\textwidth}    
     \centerline{\Large \bf     
      \hspace{0.38 \textwidth}  \color{black}{(c)}
      \hspace{0.42\textwidth}  \color{black}{(d)}
         \hfill}
     \vspace{0.29\textwidth}    
    \caption[]{Intensity (a), emission measure (b), temperature (c), and electron density (d) of an HXP of the second group (28 February 2002).}
    \label{F:T_I_n_EM_2nd}
\end{figure}

Let us describe the behavior of the second group (see Figure~\ref{F:T_I_n_EM_2nd}). HXPs of this group have an abrupt jump in temperature to high values (30\todash 50~MK) over a short period of time (2\todash 5~minutes) with slowly changing intensity. The HXP, after reaching its maximum temperature,  cools down rapidly to 10\todash 20~MK in 2\todash 5~minutes. After this, the temperature stops changing and remains constant during the latter part of its lifetime (10\todash 30~min). Intensity and emission measure reach their maxima when the temperature remains constant. After reaching its maximum intensity and maximum emission measure, intensity and emission measure simultaneously  decrease slowly to zero. Since the cadence of the spectroheliograph  is limited, the  temperature increase phase is not seen. That is why some events are seen beginning from stationary temperatures (10\todash 20~MK). The HXPs of the first group disappear  because their temperatures leave the range, which can be seen using a spectroheliograph. The HXPs of the second group disappear because their emission measures approach zero. A decrease in the emission measure could take place because of the expanding (increasing volume) of the HXP or because of a decrease of electron density (or  both).
HXPs of the second group represent 40\% of the whole observed HXPs.

In addition to  these two groups, there were very short events with  lifetimes of two\todash three minutes. These events were seen during only two\todash four frames, and therefore very little can be said about their variation.
Fast HXPs are 20\% of the  observed HXPs.

\section{Spatial Distribution of HXP}
\begin{figure}
\centerline{\includegraphics[width = \textwidth, clip=]{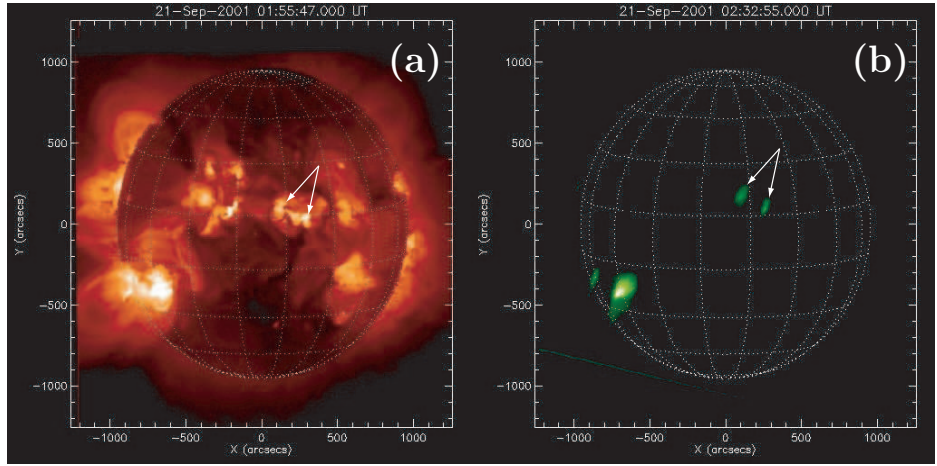}
}
\vspace{-0.46\textwidth}   
     \centerline{\Large \bf     
      \hspace{0.39 \textwidth}  \color{white}{(a)}
      \hspace{0.41\textwidth}  \color{white}{(b)}
         \hfill}
     \vspace{0.42\textwidth}    

    \caption[]{\textit{Yohkoh}/SXT (a) and \ion{Mg}{xii} spectroheliograph (b). HXPs, which are loop footpoints,  are marked with arrows.}
    \label{F:loop_footpoint}
\end{figure}
In order to investigate possible connection between HXPs and active regions, the  \ion{Mg}{xii} spectroheliograph  compared with  \textit{Yohkoh}/SXT images. \textit{Yohkoh}/SXT has a wide temperature range ($T>$~2~MK), which includes hot plasma ($T>$~5~MK, which is seen in the \ion{Mg}{xii} spectroheliograph) and cool components ($T$~=~2\todash 5~MK, in which active regions are seen). Between  August and December 2001 both satellites were operating, and a comparison could be carried out  at this period of time.

Examples of such frames are shown in Figure~\ref{F:compact_structure} and Figure~\ref{F:loop_footpoint}. In Figure~\ref{F:compact_structure} images of \textit{Yohkoh}/SXT and spectroheliograph taken within one hour are shown. HXPs,  which are compact structures, are marked with  arrows 1\todash 3. In Figure~\ref{F:loop_footpoint} images of \textit{Yohkoh}/SXT and spectroheliograph taken at close points in time  are shown. HXPs,  which are loop footpoints, are marked with  arrows. So HXPs could be a compact structure or be a part of an active region (footpoints of its loops).

\begin{figure}
\centerline{\includegraphics[width = 0.6\textwidth, clip=]{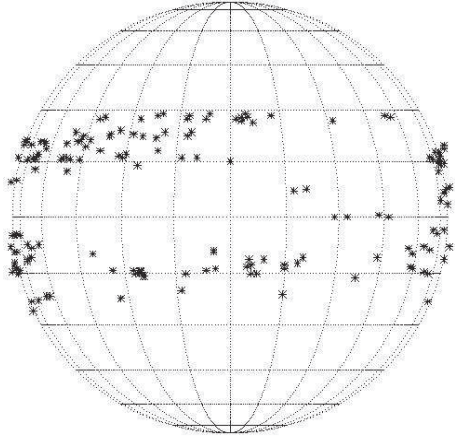}
}
    \caption[]{Spatial distribution of HXPs}
    \label{F:Space_distribution}
\end{figure}

In Figure~\ref{F:Space_distribution} the spatial distribution of HXPs over the Sun's surface is shown. In order to build this map, coordinates of 169 HXPs observed between 2002  20 and 28 February were used. Figure~\ref{F:Space_distribution} shows that HXPs are concentrated in the active-region bands. Microflares are also concentrated in the active region bands \cite{Christe2008}. X-ray bright points (XBP) are uniformly spread over the solar surface \cite{gol74}. Thus, HXPs have a similar spatial distribution with microflares but are different than XBPs.

\section{Cooling Time and Energy of HXP}
Comparison of heating and cooling times of HXPs with conductive and radiative cooling rate will allow us to study the heating mechanism. A typical value of a HXP heating time is five minutes; the cooling time lies in the range of 2\todash 30~minutes. Let us compare these values with radiative and conductive cooling times.

The conductive cooling time $\tau_\mathrm{cond}$ is  \cite{cul94}
\begin{equation}
\tau_\mathrm{cond} = \frac{21 n_e k_B h^2}{5 \kappa T^{5/2}} =
\frac{21 k_B \sqrt{EM \cdot h}}{5 \kappa T^{5/2}} \le
\frac{21 k_B \sqrt{EM \cdot h_0}}{5 \kappa T^{5/2}} = 15 \ \textmd{s}
\end{equation}
(where $\kappa = 9.2 \times 10^{-7} \ \textmd{erg} \ \textmd{s}^{-1}  \textmd{cm}^{-1}  \textmd{K}^{-7/2}$ is the Spitzer conductivity, $EM \approx 10^{47} \ \textmd{cm}^{-3}$ is the HXP emission measure, $T \approx 10^7 \ \textmd{K}$ is the HXP temperature, $h$ is the source size, $h_0$ is the effective pixel size).

Let us estimate the influence of radiative losses. The radiative loss rate is
\begin{equation}
P_\mathrm{rad} = EM \ \Lambda (T),
\end{equation}
where $\Lambda (T) \approx 10^{-22} \ \textmd{erg} \ \textmd{cm}^3 \ \textmd{s}^{-1}$ is the radiative loss function \cite{Klimchuk08}.

The conductive loss rate is
\begin{equation}
P_\mathrm{cond} = \kappa S \frac{\partial T}{\partial x} \sim
\kappa T^{\frac52}h^2\frac{T}{h} = \kappa T^{\frac72}h
\end{equation}

Let us compare these two quantities:
\begin{equation}
\frac{P_\mathrm{cond}}{P_\mathrm{rad}} \sim \frac{\kappa T^{\frac72}h}{EM \cdot \Lambda (T)} \sim 10^2
\end{equation}
so radiative losses  can be neglected.

The conductive cooling time is significantly lower than the HXP lifetime, heating time, and cooling time. This means that energy release in the HXP volume occurs throughout  its lifetime.

HXP thermal energy $\varepsilon_\mathrm{th}$ can be estimated as
\begin{equation}
\varepsilon_\mathrm{th} \sim n V k_B T =
k_B T \sqrt{EM \ V} \le
k_B T \sqrt{EM \ h_0^3} = 10^{28} \ \textmd{erg}
\end{equation}
$\varepsilon_\mathrm{th}$ is four orders lower than thermal energy of the most powerful flares  ($4 \times 10^{32}$~erg: \inlinecite{hud83}). This means that HXPs could be classified by thermal energy scale as a micro-scale events.

Since HXP lifetime is significantly greater than $\tau_\mathrm{cond}$, the power, which is heating the HXP, must be approximately equal to conductive losses:
\begin{equation}
P \approx \frac{\varepsilon_\mathrm{th}}{\tau_\mathrm{cond}}\approx 10^{27}\ \textmd{erg\ s$^{-1}$}
\end{equation}

The total energy, which is released inside HXP, equals
\begin{equation}
\varepsilon_\mathrm{total} \approx P \ \tau_\mathrm{total} \approx 10^{30}\ \textmd{erg},
\end{equation}
where $\tau_\mathrm{total}$ is the HXP lifetime.

\section{HXPs and Other Microactivity Phenomena}
Let us compare  measured HXP physical parameters with parameters of other flare-like microevents: microflares \cite{Christe2008,Hannah2008}, XBPs \cite{gol76,golub97} and nanoflares \cite{Par88,asc00}. The parameters of these phenomena are listed in Table~\ref{L:events}.
\begin{table}[h]
\caption{Comparison of parameters of HXPs and other microactivity phenomena}
\label{L:events}
\begin{tabular}{lcccc}
\hline
            & HXP           & XBP        & microflares   & nanoflares \\
\hline
Lifetime    & 5\todash 100 min & 8\todash 40 h & 1\todash 10 min & 1\todash 3 hr \\
$T$[MK]     & 5\todash 50      & 1\todash 2    & 10\todash 15    & 1\todash 2 \\
$EM$[\cmt]  & $10^{45}$\todash $10^{48}$  & $\approx 10^{47}$& $10^{45}$\todash $10^{48}$ & $\approx 10^{44}$ \\
$n_e$[\cmt] & $\ge 10^{10}$         & $\approx 5\cdot10^9$& $10^9$\todash $10^{11}$ & $10^8$\todash $10^9$ \\
Size[Mm]  & $\le 5$               & $\approx 25$        & 5\todash 25 & 2\todash 20 \\
\hline
\end{tabular}
\end{table}

Table~\ref{L:events} shows that HXPs  differ significantly from XBPs and nanoflares in their lifetimes, temperatures, and sizes. HXPs have higher emission measures and electron densities than nanoflares. HXPs differ from XBPs in their spatial distribution over the solar surface.

HXPs and microflares have the same range of values of emission measure, electron density, and thermal energy. Also they have the same spatial distribution over the solar surface (they are concentrated in the active-region latitudes). Nonetheless, HXPs have longer lifetimes, higher temperatures, and smaller sizes than microflares.

Taking into account these arguments, we  conclude that HXPs, nanoflares, and XBPs are different phenomena. The differences between HXPs and microflares are less clear, and we cannot rule out the possibility  that they belong to a similar class of phenomena.
\appendix
\section{Influence of Turbulence}
\label{A:turb}

Turbulence contributes to \ion{Mg}{xii} ion line broadening:
\begin{equation}
\left( \frac{\Delta \lambda}{\lambda} \right)^2 =
 \frac{k_B T}{M}+u^2_\mathrm{turb},
\label{E:Doppler_turb}
\end{equation}
where $M$ is the mass of magnesium ion, $u_\mathrm{turb}$ is the turbulence velocity. Equation~(\ref{E:Doppler_turb}) shows that turbulence increases the value of the effective temperature, obtained by our method, by the amount:
\begin{equation}
\Delta T_\mathrm{turb} = \frac{M u_\mathrm{turb}^2}{k_\mathrm{B}}
\end{equation}
The uncertainty in the temperature  in our work is $\Delta T_\mathrm{err} \approx$~1--2~MK. If $u_\mathrm{turb} =$~30~km~s$^{-1}$, then $\Delta T_\mathrm{turb}$ will be equal to $\Delta T_\mathrm{err}$. That means that only high values of turbulence will affect the value of the  temperature obtained. If $u_\mathrm{turb}$ is less than or equal to 30~km~s$^{-1}$ then the  $\Delta T_\mathrm{turb}$ will be less than the error of our method. That is why we neglect turbulence in our work.

Figure~\ref{F:Temperature_hist} shows a histogram of HXP peak temperatures. It has a maximum around 12~MK and a high-temperature flat tail, which starts at 20~MK and ends at 50~MK. This tail could be caused by nonthermal line broadening due to turbulence. If we suppose that the ``true'' temperatures of HXPs of this tail are around 10~MK, then the $\Delta T_\mathrm{turb}$ will be 30~MK, which would arise from $u_\mathrm{turb} \approx$~180~km~s$^{-1}$. This is a very high value, but nonetheless possible. Such values are encountered in large solar flares \cite{Kay2006}.

High observed temperatures could indeed be caused by nonthermal broadening due to turbulence. But such an explanation requires an assumption of high turbulence velocities.
\begin{acks}
The authors acknowledge the help of A.M. Urnov, N.K. Suhodrev,  and J. Sylwester  with preparing of this article for publication. The work was supported by the Russian Foundation of Basic Research (grant  11-02-01079a), by a grant of the     President of Russian Federation (MK-3875.2011.2), and the SOTERIA project of Programme FP7/2007-2013 (grant 218816).
\end{acks}

\bibliographystyle{spr-mp-sola}
\bibliography{mybibl}

\begin{thebibliography}{23}
\ifx \bisbn   \undefined \def \bisbn  #1{ISBN #1}\fi
\ifx \binits  \undefined \def \binits#1{#1}\fi
\ifx \bauthor  \undefined \def \bauthor#1{#1}\fi
\ifx \batitle  \undefined \def \batitle#1{#1}\fi
\ifx \bjtitle  \undefined \def \bjtitle#1{\textit{#1}}\fi
\ifx \bvolume  \undefined \def \bvolume#1{\textbf{#1}}\fi
\ifx \byear  \undefined \def \byear#1{#1}\fi
\ifx \bissue  \undefined \def \bissue#1{#1}\fi
\ifx \bfpage  \undefined \def \bfpage#1{#1}\fi
\ifx \blpage  \undefined \def \blpage #1{#1}\fi
\ifx \burl  \undefined \def \burl#1{\textsf{#1}}\fi
\ifx \href  \undefined \def \href#1#2{\textsf{#2}}\fi
\ifx \doiurl  \undefined \def
  \doiurl#1{\href{http://dx.doi.org/#1}{\textsf{#1}}}\fi
\ifx \betal  \undefined \def \betal{\textit{et al.}}\fi
\ifx \binstitute  \undefined \def \binstitute#1{#1}\fi
\ifx \bctitle  \undefined \def \bctitle#1{#1}\fi
\ifx \beditor  \undefined \def \beditor#1{#1}\fi
\ifx \bpublisher  \undefined \def \bpublisher#1{#1}\fi
\ifx \bbtitle  \undefined \def \bbtitle#1{\textit{#1}}\fi
\ifx \bedition  \undefined \def \bedition#1{#1}\fi
\ifx \bseriesno  \undefined \def \bseriesno#1{\textbf{#1}}\fi
\ifx \blocation  \undefined \def \blocation#1{#1}\fi
\ifx \bsertitle  \undefined \def \bsertitle#1{\textit{#1}}\fi
\ifx \bsnm \undefined \def \bsnm#1{#1}\fi
\ifx \bsuffix \undefined \def \bsuffix#1{#1}\fi
\ifx \bparticle \undefined \def \bparticle#1{#1}\fi
\ifx \barticle \undefined \def \barticle#1{}\fi
\ifx \botherref \undefined \def \botherref#1{}\fi
\ifx \url \undefined \def \url#1{\textsf{#1}}\fi
\ifx \bchapter \undefined \def \bchapter#1{}\fi
\ifx \bbook \undefined \def \bbook#1{}\fi
\ifx \bcomment \undefined \def \bcomment#1{#1}\fi
\ifx \oauthor \undefined \def \oauthor#1{#1}\fi
\ifx \citeauthoryear \undefined \def \citeauthoryear#1{#1}\fi
\def \endbibitem {}
\ifx \bconflocation  \undefined \def \bconflocation#1{#1} \fi

\bibitem[\protect\citeauthoryear{{Aschwanden} \textit{et~al.}}{2000}]{asc00}
\begin{barticle}
\bauthor{\bsnm{{Aschwanden}}, \binits{M.J.}},
\bauthor{\bsnm{{Tarbell}}, \binits{T.D.}},
\bauthor{\bsnm{{Nightingale}}, \binits{R.W.}},
\bauthor{\bsnm{{Schrijver}}, \binits{C.J.}},
\bauthor{\bsnm{{Title}}, \binits{A.}},
\bauthor{\bsnm{{Kankelborg}}, \binits{C.C.}},
\bauthor{\bsnm{{Martens}}, \binits{P.}},
\bauthor{\bsnm{{Warren}}, \binits{H.P.}}:
\byear{2000},
\batitle{{Time Variability of the ``Quiet'' Sun Observed with TRACE. II.
  Physical Parameters, Temperature Evolution, and Energetics of
  Extreme-Ultraviolet Nanoflares}}.
\bjtitle{\apj}
\bvolume{535},
\bfpage{1047}\,--\,\blpage{1065}.
doi:\doiurl{10.1086/308867}.
\end{barticle}
\endbibitem

\bibitem[\protect\citeauthoryear{{Benz} and {Grigis}}{2002}]{ben02}
\begin{barticle}
\bauthor{\bsnm{{Benz}}, \binits{A.O.}},
\bauthor{\bsnm{{Grigis}}, \binits{P.C.}}:
\byear{2002},
\batitle{{Microflares and hot component in solar active regions}}.
\bjtitle{\solphys}
\bvolume{210},
\bfpage{431}\,--\,\blpage{444}.
doi:\doiurl{10.1023/A:1022496515506}.
\end{barticle}
\endbibitem

\bibitem[\protect\citeauthoryear{{Christe} \textit{et~al.}}{2008}]{Christe2008}
\begin{barticle}
\bauthor{\bsnm{{Christe}}, \binits{S.}},
\bauthor{\bsnm{{Hannah}}, \binits{I.G.}},
\bauthor{\bsnm{{Krucker}}, \binits{S.}},
\bauthor{\bsnm{{McTiernan}}, \binits{J.}},
\bauthor{\bsnm{{Lin}}, \binits{R.P.}}:
\byear{2008},
\batitle{{RHESSI Microflare Statistics. I. Flare-Finding and Frequency
  Distributions}}.
\bjtitle{\apj}
\bvolume{677},
\bfpage{1385}\,--\,\blpage{1394}.
doi:\doiurl{10.1086/529011}.
\end{barticle}
\endbibitem

\bibitem[\protect\citeauthoryear{{Culhane} \textit{et~al.}}{1994}]{cul94}
\begin{barticle}
\bauthor{\bsnm{{Culhane}}, \binits{J.L.}},
\bauthor{\bsnm{{Phillips}}, \binits{A.T.}},
\bauthor{\bsnm{{Inda-Koide}}, \binits{M.}},
\bauthor{\bsnm{{Kosugi}}, \binits{T.}},
\bauthor{\bsnm{{Fludra}}, \binits{A.}},
\bauthor{\bsnm{{Kurokawa}}, \binits{H.}},
\bauthor{\bsnm{{Makishima}}, \binits{K.}},
\bauthor{\bsnm{{Pike}}, \binits{C.D.}},
\bauthor{\bsnm{{Sakao}}, \binits{T.}},
\bauthor{\bsnm{{Sakurai}}, \binits{T.}}:
\byear{1994},
\batitle{{YOHKOH observations of the creation of high-temperature plasma in the
  flare of 16 December 1991}}.
\bjtitle{\solphys}
\bvolume{153},
\bfpage{307}\,--\,\blpage{336}.
doi:\doiurl{10.1007/BF00712508}.
\end{barticle}
\endbibitem

\bibitem[\protect\citeauthoryear{{Golub} and {Pasachoff}}{1997}]{golub97}
\begin{bbook}
\bauthor{\bsnm{{Golub}}, \binits{L.}},
\bauthor{\bsnm{{Pasachoff}}, \binits{J.M.}}:
\byear{1997},
\bbtitle{{The Solar Corona}},
\bpublisher{Cambridge University Press},
\blocation{Cambridge, UK}.
\end{bbook}
\endbibitem

\bibitem[\protect\citeauthoryear{{Golub}, {Krieger}, and
  {Vaiana}}{1976}]{gol76}
\begin{barticle}
\bauthor{\bsnm{{Golub}}, \binits{L.}},
\bauthor{\bsnm{{Krieger}}, \binits{A.S.}},
\bauthor{\bsnm{{Vaiana}}, \binits{G.S.}}:
\byear{1976},
\batitle{{Observation of spatial and temporal variations in X-ray bright point
  emergence patterns}}.
\bjtitle{\solphys}
\bvolume{50},
\bfpage{311}\,--\,\blpage{327}.
doi:\doiurl{10.1007/BF00155294}.
\end{barticle}
\endbibitem

\bibitem[\protect\citeauthoryear{{Golub} \textit{et~al.}}{1974}]{gol74}
\begin{barticle}
\bauthor{\bsnm{{Golub}}, \binits{L.}},
\bauthor{\bsnm{{Krieger}}, \binits{A.S.}},
\bauthor{\bsnm{{Silk}}, \binits{J.K.}},
\bauthor{\bsnm{{Timothy}}, \binits{A.F.}},
\bauthor{\bsnm{{Vaiana}}, \binits{G.S.}}:
\byear{1974},
\batitle{{Solar X-Ray Bright Points}}.
\bjtitle{\apjl}
\bvolume{189},
\bfpage{L93}.
doi:\doiurl{10.1086/181472}.
\end{barticle}
\endbibitem

\bibitem[\protect\citeauthoryear{{Golub} \textit{et~al.}}{2007}]{Golub2007}
\begin{barticle}
\bauthor{\bsnm{{Golub}}, \binits{L.}},
\bauthor{\bsnm{{Deluca}}, \binits{E.}},
\bauthor{\bsnm{{Austin}}, \binits{G.}},
\bauthor{\bsnm{{Bookbinder}}, \binits{J.}},
\bauthor{\bsnm{{Caldwell}}, \binits{D.}},
\bauthor{\bsnm{{Cheimets}}, \binits{P.}},
\bauthor{\bsnm{{Cirtain}}, \binits{J.}},
\bauthor{\bsnm{{Cosmo}}, \binits{M.}},
\bauthor{\bsnm{{Reid}}, \binits{P.}},
\bauthor{\bsnm{{Sette}}, \binits{A.}},
\bauthor{\bsnm{{Weber}}, \binits{M.}},
\bauthor{\bsnm{{Sakao}}, \binits{T.}},
\bauthor{\bsnm{{Kano}}, \binits{R.}},
\bauthor{\bsnm{{Shibasaki}}, \binits{K.}},
\bauthor{\bsnm{{Hara}}, \binits{H.}},
\bauthor{\bsnm{{Tsuneta}}, \binits{S.}},
\bauthor{\bsnm{{Kumagai}}, \binits{K.}},
\bauthor{\bsnm{{Tamura}}, \binits{T.}},
\bauthor{\bsnm{{Shimojo}}, \binits{M.}},
\bauthor{\bsnm{{McCracken}}, \binits{J.}},
\bauthor{\bsnm{{Carpenter}}, \binits{J.}},
\bauthor{\bsnm{{Haight}}, \binits{H.}},
\bauthor{\bsnm{{Siler}}, \binits{R.}},
\bauthor{\bsnm{{Wright}}, \binits{E.}},
\bauthor{\bsnm{{Tucker}}, \binits{J.}},
\bauthor{\bsnm{{Rutledge}}, \binits{H.}},
\bauthor{\bsnm{{Barbera}}, \binits{M.}},
\bauthor{\bsnm{{Peres}}, \binits{G.}},
\bauthor{\bsnm{{Varisco}}, \binits{S.}}:
\byear{2007},
\batitle{{The X-Ray Telescope (XRT) for the Hinode Mission}}.
\bjtitle{\solphys}
\bvolume{243},
\bfpage{63}\,--\,\blpage{86}.
doi:\doiurl{10.1007/s11207-007-0182-1}.
\end{barticle}
\endbibitem

\bibitem[\protect\citeauthoryear{{Hannah} \textit{et~al.}}{2008}]{Hannah2008}
\begin{barticle}
\bauthor{\bsnm{{Hannah}}, \binits{I.G.}},
\bauthor{\bsnm{{Christe}}, \binits{S.}},
\bauthor{\bsnm{{Krucker}}, \binits{S.}},
\bauthor{\bsnm{{Hurford}}, \binits{G.J.}},
\bauthor{\bsnm{{Hudson}}, \binits{H.S.}},
\bauthor{\bsnm{{Lin}}, \binits{R.P.}}:
\byear{2008},
\batitle{{RHESSI Microflare Statistics. II. X-Ray Imaging, Spectroscopy, and
  Energy Distributions}}.
\bjtitle{\apj}
\bvolume{677},
\bfpage{704}\,--\,\blpage{718}.
doi:\doiurl{10.1086/529012}.
\end{barticle}
\endbibitem

\bibitem[\protect\citeauthoryear{{Hudson} and {Willson}}{1983}]{hud83}
\begin{barticle}
\bauthor{\bsnm{{Hudson}}, \binits{H.S.}},
\bauthor{\bsnm{{Willson}}, \binits{R.C.}}:
\byear{1983},
\batitle{{Upper limits on the total radiant energy of solar flares}}.
\bjtitle{\solphys}
\bvolume{86},
\bfpage{123}\,--\,\blpage{130}.
doi:\doiurl{10.1007/BF00157181}.
\end{barticle}
\endbibitem

\bibitem[\protect\citeauthoryear{{Kay} \textit{et~al.}}{2006}]{Kay2006}
\begin{barticle}
\bauthor{\bsnm{{Kay}}, \binits{H.R.M.}},
\bauthor{\bsnm{{Matthews}}, \binits{S.A.}},
\bauthor{\bsnm{{Harra}}, \binits{L.K.}},
\bauthor{\bsnm{{Culhane}}, \binits{J.L.}}:
\byear{2006},
\batitle{{Non-thermal broadening of coronal emission lines in the onset phase
  of solar flares and CMEs}}.
\bjtitle{\aap}
\bvolume{447},
\bfpage{719}\,--\,\blpage{725}.
doi:\doiurl{10.1051/0004-6361:20053240}.
\end{barticle}
\endbibitem

\bibitem[\protect\citeauthoryear{{Klimchuk}, {Patsourakos}, and
  {Cargill}}{2008}]{Klimchuk08}
\begin{barticle}
\bauthor{\bsnm{{Klimchuk}}, \binits{J.A.}},
\bauthor{\bsnm{{Patsourakos}}, \binits{S.}},
\bauthor{\bsnm{{Cargill}}, \binits{P.J.}}:
\byear{2008},
\batitle{{Highly Efficient Modeling of Dynamic Coronal Loops}}.
\bjtitle{\apj}
\bvolume{682},
\bfpage{1351}\,--\,\blpage{1362}.
doi:\doiurl{10.1086/589426}.
\end{barticle}
\endbibitem

\bibitem[\protect\citeauthoryear{{Kosugi} \textit{et~al.}}{2007}]{Kosugi2007}
\begin{barticle}
\bauthor{\bsnm{{Kosugi}}, \binits{T.}},
\bauthor{\bsnm{{Matsuzaki}}, \binits{K.}},
\bauthor{\bsnm{{Sakao}}, \binits{T.}},
\bauthor{\bsnm{{Shimizu}}, \binits{T.}},
\bauthor{\bsnm{{Sone}}, \binits{Y.}},
\bauthor{\bsnm{{Tachikawa}}, \binits{S.}},
\bauthor{\bsnm{{Hashimoto}}, \binits{T.}},
\bauthor{\bsnm{{Minesugi}}, \binits{K.}},
\bauthor{\bsnm{{Ohnishi}}, \binits{A.}},
\bauthor{\bsnm{{Yamada}}, \binits{T.}},
\bauthor{\bsnm{{Tsuneta}}, \binits{S.}},
\bauthor{\bsnm{{Hara}}, \binits{H.}},
\bauthor{\bsnm{{Ichimoto}}, \binits{K.}},
\bauthor{\bsnm{{Suematsu}}, \binits{Y.}},
\bauthor{\bsnm{{Shimojo}}, \binits{M.}},
\bauthor{\bsnm{{Watanabe}}, \binits{T.}},
\bauthor{\bsnm{{Shimada}}, \binits{S.}},
\bauthor{\bsnm{{Davis}}, \binits{J.M.}},
\bauthor{\bsnm{{Hill}}, \binits{L.D.}},
\bauthor{\bsnm{{Owens}}, \binits{J.K.}},
\bauthor{\bsnm{{Title}}, \binits{A.M.}},
\bauthor{\bsnm{{Culhane}}, \binits{J.L.}},
\bauthor{\bsnm{{Harra}}, \binits{L.K.}},
\bauthor{\bsnm{{Doschek}}, \binits{G.A.}},
\bauthor{\bsnm{{Golub}}, \binits{L.}}:
\byear{2007},
\batitle{{The Hinode (Solar-B) Mission: An Overview}}.
\bjtitle{\solphys}
\bvolume{243},
\bfpage{3}\,--\,\blpage{17}.
doi:\doiurl{10.1007/s11207-007-9014-6}.
\end{barticle}
\endbibitem

\bibitem[\protect\citeauthoryear{{Lin} \textit{et~al.}}{1984}]{Lin1984}
\begin{barticle}
\bauthor{\bsnm{{Lin}}, \binits{R.P.}},
\bauthor{\bsnm{{Schwartz}}, \binits{R.A.}},
\bauthor{\bsnm{{Kane}}, \binits{S.R.}},
\bauthor{\bsnm{{Pelling}}, \binits{R.M.}},
\bauthor{\bsnm{{Hurley}}, \binits{K.C.}}:
\byear{1984},
\batitle{{Solar hard X-ray microflares}}.
\bjtitle{\apj}
\bvolume{283},
\bfpage{421}\,--\,\blpage{425}.
doi:\doiurl{10.1086/162321}.
\end{barticle}
\endbibitem

\bibitem[\protect\citeauthoryear{{Lin} \textit{et~al.}}{2002}]{Lin2002}
\begin{barticle}
\bauthor{\bsnm{{Lin}}, \binits{R.P.}},
\bauthor{\bsnm{{Dennis}}, \binits{B.R.}},
\bauthor{\bsnm{{Hurford}}, \binits{G.J.}},
\bauthor{\bsnm{{Smith}}, \binits{D.M.}},
\bauthor{\bsnm{{Zehnder}}, \binits{A.}},
\bauthor{\bsnm{{Harvey}}, \binits{P.R.}},
\bauthor{\bsnm{{Curtis}}, \binits{D.W.}},
\bauthor{\bsnm{{Pankow}}, \binits{D.}},
\bauthor{\bsnm{{Turin}}, \binits{P.}},
\bauthor{\bsnm{{Bester}}, \binits{M.}},
\bauthor{\bsnm{{Csillaghy}}, \binits{A.}},
\bauthor{\bsnm{{Lewis}}, \binits{M.}},
\bauthor{\bsnm{{Madden}}, \binits{N.}},
\bauthor{\bsnm{{van Beek}}, \binits{H.F.}},
\bauthor{\bsnm{{Appleby}}, \binits{M.}},
\bauthor{\bsnm{{Raudorf}}, \binits{T.}},
\bauthor{\bsnm{{McTiernan}}, \binits{J.}},
\bauthor{\bsnm{{Ramaty}}, \binits{R.}},
\bauthor{\bsnm{{Schmahl}}, \binits{E.}},
\bauthor{\bsnm{{Schwartz}}, \binits{R.}},
\bauthor{\bsnm{{Krucker}}, \binits{S.}},
\bauthor{\bsnm{{Abiad}}, \binits{R.}},
\bauthor{\bsnm{{Quinn}}, \binits{T.}},
\bauthor{\bsnm{{Berg}}, \binits{P.}},
\bauthor{\bsnm{{Hashii}}, \binits{M.}},
\bauthor{\bsnm{{Sterling}}, \binits{R.}},
\bauthor{\bsnm{{Jackson}}, \binits{R.}},
\bauthor{\bsnm{{Pratt}}, \binits{R.}},
\bauthor{\bsnm{{Campbell}}, \binits{R.D.}},
\bauthor{\bsnm{{Malone}}, \binits{D.}},
\bauthor{\bsnm{{Landis}}, \binits{D.}},
\bauthor{\bsnm{{Barrington-Leigh}}, \binits{C.P.}},
\bauthor{\bsnm{{Slassi-Sennou}}, \binits{S.}},
\bauthor{\bsnm{{Cork}}, \binits{C.}},
\bauthor{\bsnm{{Clark}}, \binits{D.}},
\bauthor{\bsnm{{Amato}}, \binits{D.}},
\bauthor{\bsnm{{Orwig}}, \binits{L.}},
\bauthor{\bsnm{{Boyle}}, \binits{R.}},
\bauthor{\bsnm{{Banks}}, \binits{I.S.}},
\bauthor{\bsnm{{Shirey}}, \binits{K.}},
\bauthor{\bsnm{{Tolbert}}, \binits{A.K.}},
\bauthor{\bsnm{{Zarro}}, \binits{D.}},
\bauthor{\bsnm{{Snow}}, \binits{F.}},
\bauthor{\bsnm{{Thomsen}}, \binits{K.}},
\bauthor{\bsnm{{Henneck}}, \binits{R.}},
\bauthor{\bsnm{{McHedlishvili}}, \binits{A.}},
\bauthor{\bsnm{{Ming}}, \binits{P.}},
\bauthor{\bsnm{{Fivian}}, \binits{M.}},
\bauthor{\bsnm{{Jordan}}, \binits{J.}},
\bauthor{\bsnm{{Wanner}}, \binits{R.}},
\bauthor{\bsnm{{Crubb}}, \binits{J.}},
\bauthor{\bsnm{{Preble}}, \binits{J.}},
\bauthor{\bsnm{{Matranga}}, \binits{M.}},
\bauthor{\bsnm{{Benz}}, \binits{A.}},
\bauthor{\bsnm{{Hudson}}, \binits{H.}},
\bauthor{\bsnm{{Canfield}}, \binits{R.C.}},
\bauthor{\bsnm{{Holman}}, \binits{G.D.}},
\bauthor{\bsnm{{Crannell}}, \binits{C.}},
\bauthor{\bsnm{{Kosugi}}, \binits{T.}},
\bauthor{\bsnm{{Emslie}}, \binits{A.G.}},
\bauthor{\bsnm{{Vilmer}}, \binits{N.}},
\bauthor{\bsnm{{Brown}}, \binits{J.C.}},
\bauthor{\bsnm{{Johns-Krull}}, \binits{C.}},
\bauthor{\bsnm{{Aschwanden}}, \binits{M.}},
\bauthor{\bsnm{{Metcalf}}, \binits{T.}},
\bauthor{\bsnm{{Conway}}, \binits{A.}}:
\byear{2002},
\batitle{{The Reuven Ramaty High-Energy Solar Spectroscopic Imager (RHESSI)}}.
\bjtitle{\solphys}
\bvolume{210},
\bfpage{3}\,--\,\blpage{32}.
doi:\doiurl{10.1023/A:1022428818870}.
\end{barticle}
\endbibitem

\bibitem[\protect\citeauthoryear{{Ogawara} \textit{et~al.}}{1991}]{Ogawara1991}
\begin{barticle}
\bauthor{\bsnm{{Ogawara}}, \binits{Y.}},
\bauthor{\bsnm{{Takano}}, \binits{T.}},
\bauthor{\bsnm{{Kato}}, \binits{T.}},
\bauthor{\bsnm{{Kosugi}}, \binits{T.}},
\bauthor{\bsnm{{Tsuneta}}, \binits{S.}},
\bauthor{\bsnm{{Watanabe}}, \binits{T.}},
\bauthor{\bsnm{{Kondo}}, \binits{I.}},
\bauthor{\bsnm{{Uchida}}, \binits{Y.}}:
\byear{1991},
\batitle{{The Solar-A Mission - an Overview}}.
\bjtitle{\solphys}
\bvolume{136},
\bfpage{1}\,--\,\blpage{16}.
doi:\doiurl{10.1007/BF00151692}.
\end{barticle}
\endbibitem

\bibitem[\protect\citeauthoryear{{Parker}}{1988}]{Par88}
\begin{barticle}
\bauthor{\bsnm{{Parker}}, \binits{E.N.}}:
\byear{1988},
\batitle{{Nanoflares and the solar X-ray corona}}.
\bjtitle{\apj}
\bvolume{330},
\bfpage{474}\,--\,\blpage{479}.
doi:\doiurl{10.1086/166485}.
\end{barticle}
\endbibitem

\bibitem[\protect\citeauthoryear{{Shimizu}}{1995}]{shi95}
\begin{barticle}
\bauthor{\bsnm{{Shimizu}}, \binits{T.}}:
\byear{1995},
\batitle{{Energetics and Occurrence Rate of Active-Region Transient
  Brightenings and Implications for the Heating of the Active-Region Corona}}.
\bjtitle{\pasj}
\bvolume{47},
\bfpage{251}\,--\,\blpage{263}.
\end{barticle}
\endbibitem

\bibitem[\protect\citeauthoryear{{Sylwester}, {Garcia}, and
  {Sylwester}}{1995}]{Sylwester1995}
\begin{barticle}
\bauthor{\bsnm{{Sylwester}}, \binits{J.}},
\bauthor{\bsnm{{Garcia}}, \binits{H.A.}},
\bauthor{\bsnm{{Sylwester}}, \binits{B.}}:
\byear{1995},
\batitle{{Quantitative interpretation of GOES soft X-ray measurements. I. The
  isothermal approximation: application of various atomic data.}}
\bjtitle{\aap}
\bvolume{293},
\bfpage{577}\,--\,\blpage{585}.
\end{barticle}
\endbibitem

\bibitem[\protect\citeauthoryear{{Tsuneta} \textit{et~al.}}{1991}]{Tsuneta1991}
\begin{barticle}
\bauthor{\bsnm{{Tsuneta}}, \binits{S.}},
\bauthor{\bsnm{{Acton}}, \binits{L.}},
\bauthor{\bsnm{{Bruner}}, \binits{M.}},
\bauthor{\bsnm{{Lemen}}, \binits{J.}},
\bauthor{\bsnm{{Brown}}, \binits{W.}},
\bauthor{\bsnm{{Caravalho}}, \binits{R.}},
\bauthor{\bsnm{{Catura}}, \binits{R.}},
\bauthor{\bsnm{{Freeland}}, \binits{S.}},
\bauthor{\bsnm{{Jurcevich}}, \binits{B.}},
\bauthor{\bsnm{{Owens}}, \binits{J.}}:
\byear{1991},
\batitle{{The soft X-ray telescope for the SOLAR-A mission}}.
\bjtitle{\solphys}
\bvolume{136},
\bfpage{37}\,--\,\blpage{67}.
doi:\doiurl{10.1007/BF00151694}.
\end{barticle}
\endbibitem

\bibitem[\protect\citeauthoryear{{Urnov} \textit{et~al.}}{2007}]{Urnov07}
\begin{barticle}
\bauthor{\bsnm{{Urnov}}, \binits{A.M.}},
\bauthor{\bsnm{{Shestov}}, \binits{S.V.}},
\bauthor{\bsnm{{Bogachev}}, \binits{S.A.}},
\bauthor{\bsnm{{Goryaev}}, \binits{F.F.}},
\bauthor{\bsnm{{Zhitnik}}, \binits{I.A.}},
\bauthor{\bsnm{{Kuzin}}, \binits{S.V.}}:
\byear{2007},
\batitle{{On the spatial and temporal characteristics and formation mechanisms
  of soft X-ray emission in the solar corona}}.
\bjtitle{\al}
\bvolume{33},
\bfpage{396}\,--\,\blpage{410}.
doi:\doiurl{10.1134/S1063773707060059}.
\end{barticle}
\endbibitem

\bibitem[\protect\citeauthoryear{{Zhitnik} \textit{et~al.}}{2003a}]{zhi03a}
\begin{barticle}
\bauthor{\bsnm{{Zhitnik}}, \binits{I.A.}},
\bauthor{\bsnm{{Bugaenko}}, \binits{O.I.}},
\bauthor{\bsnm{{Ignat'ev}}, \binits{A.P.}},
\bauthor{\bsnm{{Krutov}}, \binits{V.V.}},
\bauthor{\bsnm{{Kuzin}}, \binits{S.V.}},
\bauthor{\bsnm{{Mitrofanov}}, \binits{A.V.}},
\bauthor{\bsnm{{Oparin}}, \binits{S.N.}},
\bauthor{\bsnm{{Pertsov}}, \binits{A.A.}},
\bauthor{\bsnm{{Slemzin}}, \binits{V.A.}},
\bauthor{\bsnm{{Stepanov}}, \binits{A.I.}},
\bauthor{\bsnm{{Urnov}}, \binits{A.M.}}:
\byear{2003}a,
\batitle{{Dynamic 10 MK plasma structures observed in monochromatic full-Sun
  images by the SPIRIT spectroheliograph on the CORONAS-F mission}}.
\bjtitle{\mnras}
\bvolume{338},
\bfpage{67}\,--\,\blpage{71}.
doi:\doiurl{10.1046/j.1365-8711.2003.06014.x}.
\end{barticle}
\endbibitem

\bibitem[\protect\citeauthoryear{{Zhitnik} \textit{et~al.}}{2003b}]{zhi03b}
\begin{barticle}
\bauthor{\bsnm{{Zhitnik}}, \binits{I.}},
\bauthor{\bsnm{{Kuzin}}, \binits{S.}},
\bauthor{\bsnm{{Afanas'ev}}, \binits{A.}},
\bauthor{\bsnm{{Bugaenko}}, \binits{O.}},
\bauthor{\bsnm{{Ignat'ev}}, \binits{A.}},
\bauthor{\bsnm{{Krutov}}, \binits{V.}},
\bauthor{\bsnm{{Mitrofanov}}, \binits{A.}},
\bauthor{\bsnm{{Oparin}}, \binits{S.}},
\bauthor{\bsnm{{Pertsov}}, \binits{A.}},
\bauthor{\bsnm{{Slemzin}}, \binits{V.}},
\bauthor{\bsnm{{Sukhodrev}}, \binits{N.}},
\bauthor{\bsnm{{Umov}}, \binits{A.}}:
\byear{2003}b,
\batitle{{XUV observations of solar corona in the spirit experiment on board
  the coronas-F satellite}}.
\bjtitle{\adv}
\bvolume{32},
\bfpage{473}\,--\,\blpage{477}.
doi:\doiurl{10.1016/S0273-1177(03)00351-X}.
\end{barticle}
\endbibitem

\end{thebibliography}

\end{article}
\end{document}